\documentclass[pra,aps,twocolumn,floatfix,showpacs]{revtex4-1}
\usepackage{graphicx,amsmath,amssymb,times}

\begin{document}
\title{Cooper pairing and BCS-BEC evolution in mixed-dimensional Fermi gases}
\author{M. Iskin}
\affiliation{Department of Physics, Ko\c c University, Rumelifeneri Yolu, 34450 Sariyer, Istanbul, Turkey.}
\author{A. L. Suba{\c s}{\i}}
\altaffiliation{Current address: 47'nci Muht. Bl. K.ligi PK:75 Selcuklu, Konya, Turkey.}
\affiliation{Faculty of Engineering and Natural Sciences, Sabanci University, 34956 Tuzla, Istanbul, Turkey.}
\date{\today}

\begin{abstract}

Similar to what has recently been achieved with Bose-Bose 
mixtures~\cite{lamporesi}, mixed-dimensional Fermi-Fermi mixtures 
can be created by applying a species-selective one-dimensional 
optical lattice to a two-species Fermi gas 
($\sigma \equiv \{ \uparrow, \downarrow \}$), such a way that 
both species are confined to quasi-two-dimensional geometries
determined by their hoppings along the lattice direction. 
We investigate the ground state 
phase diagram of superfluidity for such mixtures in the BCS-BEC evolution,
and find normal, gapped superfluid, gapless superfluid, 
and phase separated regions. In particular, we find a stable gapless 
superfluid phase where the unpaired $\uparrow$ and $\downarrow$ 
fermions coexist with the paired (or superfluid) ones 
in different momentum space regions. This phase is in some ways 
similar to the Sarma state found in mixtures with unequal densities,
but in our case, the gapless superfluid phase is unpolarized and 
most importantly it is stable against phase separation.

\end{abstract}

\maketitle

\section{Introduction}
\label{sec:introduction}

Atomic Fermi gases have emerged as unique testing ground for many 
theories of exotic matter in nature, allowing for the creation of 
complex yet very controllable many-body quantum systems~\cite{review}, 
where for instance observation of the BCS-BEC crossover 
has so far been the most important achievement in this field.
Following this huge success with single-species fermion mixtures, 
there has been increasing experimental interest in studying 
two-species Fermi-Fermi 
mixtures~\cite{taglieber08, wille08, voigt09, spiegelhalder09, tiecke09, spiegelhalder10}.
In particular, $^6$Li-$^{40}$K mixtures have recently been 
trapped and interspecies Feshbach resonances have been identified, 
opening a new frontier in ultracold atom research to study 
exotic many-body phenomena, one of which is the possibility of studying 
fermion pairing in mixed dimensions~\cite{nishida08, nishida09, nishida10}.

Mixed-dimensional atomic systems, in which two types of particles 
live in different dimensions, can be created with two-species 
Fermi-Fermi, Bose-Fermi, and Bose-Bose mixtures by using
species-selective optical lattice potentials. 
This has recently been achieved with Bose-Bose 
mixtures~\cite{lamporesi} by applying a one-dimensional optical 
lattice to the $^{41}$K-$^{87}$Rb mixture, where only $^{41}$K 
atoms feel the lattice potential, and they are confined to a 
quasi-two-dimensional geometry, while having negligible 
effect on $^{87}$Rb atoms, that is leaving $^{87}$Rb atoms 
three dimensional. Motivated by this experimental work, 
here we analyze Cooper pairing in mixed-dimensional Fermi gases.
We consider both single-species and 
two-species fermion mixtures, and analyze the ground state phase 
diagrams in the BCS-BEC evolution, which involves normal, 
gapped superfluid, gapless superfluid, and phase separated regions.
In particular, the gapless superfluid phase, where the unpaired 
$\uparrow$ and $\downarrow$ fermions coexist 
with the paired (or superfluid) ones in different 
momentum space regions, is in some ways similar 
to the Sarma state found in mixtures with unequal densities~\cite{sarma}, 
but in our case, the gapless superfluid phase is unpolarized and 
most importantly it is stable against phase separation. 
In this way, our gapless superfluid phase is very similar to those 
of Refs.~\cite{eite, feigun}, which are recently proposed for 
ultracold atomic systems in other contexts. 

The rest of the manuscript is organized as follows. In Sec.~\ref{sec:mfg}, 
after introducing the Hamiltonian in Sec.~\ref{sec:hamiltonian},
the corresponding saddle-point self-consistency equations are
derived in Sec.~\ref{sec:self-consistency}, and their noninteracting 
limit is discussed in Sec.~\ref{sec:nil}. We numerically solve 
these equations in the BCS-BEC evolution in Sec.~\ref{sec:sps}, 
where we investigate the normal-superfluid transition in Sec.~\ref{sec:nst},
the topological gapless-gapped superfluidity transition in 
Sec.~\ref{sec:topological}, and the ground state phase diagrams in 
Sec.~\ref{sec:gspd}. A brief summary of our conclusions is given in 
Sec.~\ref{sec:conclusions}. We also include three appendices, where
the self-consistency equations are further discussed in Appendix A, 
boundary equation for the normal-superfluid transition is derived 
in Appendix B, and the molecular BEC limit is investigated in Appendix C.

\section{Mixed-dimensional Fermi gases}
\label{sec:mfg}

In this work, we analyze Cooper pairing in mixed-dimensional Fermi 
gases, which seems to be a very promising way to create superfluidity 
with mismatched Fermi surfaces, and the physics involved is in some 
ways similar to that of the unequal density 
mixtures~\cite{zwierlerin06, partridge06, shin08, salomon10}. 
We consider only uniform (homogenous) mixtures, but emphasize that 
the finite-size effects due to the confining trapping potentials 
(which are always present in atomic systems) can be taken into account 
using the local-density approximation (as a first approximation).

\subsection{Hamiltonian}
\label{sec:hamiltonian}

To describe such mixed-dimensional Fermi gases in a species-selective 
one-dimensional optical lattice (say in the $\mathbf{\widehat{z}}$ 
direction), we start with the real-space Hamiltonian ($\hbar = k_B = 1$)
\begin{align}
H &= \sum_{\sigma} \int d^3\mathbf{r} \psi_{\mathbf{r},\sigma}^\dagger 
  \left[ -\frac{\nabla^2}{2m_\sigma} -\mu_\sigma + V_{OL}^\sigma(r_z) \right] 
   \psi_{\mathbf{r},\sigma} \nonumber \\
  &- g \int d^3\mathbf{r} \psi_{\mathbf{r},\uparrow}^\dagger \psi_{\mathbf{r},\downarrow}^\dagger 
   \psi_{\mathbf{r},\downarrow} \psi_{\mathbf{r},\uparrow},
\label{eqn:realham}
\end{align}
where the pseudo-spin $\sigma$ labels both the type and hyperfine states of atoms
represented by the creation operator $\psi_{\mathbf{r},\sigma}^\dagger$,
and $m_\sigma$ is the mass and $\mu_\sigma$ is the chemical potential. Here, 
$\mathbf{r} \equiv (r_x,r_y,r_z)$ is the position with $r_\perp = \sqrt{r_x^2 + r_y^2}$,
$V_{OL}^\sigma = V_{0,\sigma} \sin^2(\pi r_z/d_z)$ is the optical lattice 
potential, $d_z$ is the lattice spacing, and $g \ge 0$ is the strength of 
the attractive interaction (zero-range and isotropic) between $\uparrow$ 
and $\downarrow$ fermions. 

In order to achieve the momentum-space Hamiltonian, we first expand the 
creation and annihilation field operators in the orthonormal and complete 
basis set of Wannier functions $W(r_z - r_z^i)$ of the lowest-energy states 
of the optical potential near their minima (single-band approximation), 
and then take the Fourier transform of the site operators. 
This corresponds to the following substitution:
$
\psi_{\mathbf{r},\sigma} = (1/\sqrt{M}) \sum_{i, \mathbf{k}} 
a_{\mathbf{k}, \sigma} W_\sigma(r_z-r_z^i) e^{-i(\mathbf{k_\perp} \cdot \mathbf{r_\perp} + k_z r_z^i)},
$
where $\mathbf{k} \equiv (k_x, k_y, k_z)$ is the momentum with $k_\perp = \sqrt{k_x^2 + k_y^2}$, 
$i$ labels the lattice sites, and $M$ is the total number of them.
Keeping only the tunneling between nearest-neighbor sites and onsite 
interactions, and using the orthonormality of the Wannier functions,
the resultant Hamiltonian can be written as
\begin{equation}
H = \sum_{\mathbf{k},\sigma} \xi_{\mathbf{k}, \sigma} a_{\mathbf{k},\sigma}^\dagger a_{\mathbf{k},\sigma} 
- g\sum_{\mathbf{k},\mathbf{k'},\mathbf{q}} b_{\mathbf{k},\mathbf{q}}^\dagger b_{\mathbf{k'},\mathbf{q}}, 
\label{eqn:hamiltonian}
\end{equation}
where 
$
b_{\mathbf{k},\mathbf{q}}^\dagger = a_{\mathbf{k}+\mathbf{q}/2,\uparrow}^\dagger 
a_{-\mathbf{k}+\mathbf{q}/2,\downarrow}^\dagger
$ 
creates fermion pairs with center of mass momentum $\mathbf{q}$ and relative 
momentum $2\mathbf{k}$. Here, 
$
\xi_{\mathbf{k},\sigma} = \epsilon_{\mathbf{k},\sigma} - \mu_\sigma,
$ 
where 
$
\epsilon_{\mathbf{k},\sigma} = k_\perp^2/(2m_\sigma) + 2t_\sigma \left[1 - \cos(k_z d_z)\right]
$
is the single-particle energy dispersion, and
$
t_\sigma = \int dr_z W^*(r_z-r_z^i)   \left[ -\partial^2/(2m_\sigma \partial r_z^2) + V_{OL}^\sigma(r_z) \right]  W(r_z-r_z^j)
$
is the tunneling amplitude between any nearest-neighbor sites $i$ and $j$. 

Following the usual treatment, strength of the attractive interaction 
can be written in terms of an ``effective'' $s$-wave scattering 
length $a_{eff}$ as
$
1/g = - m_+ V / (4\pi a_{eff}) + \sum_{\mathbf{k}} 1/(2\epsilon_{\mathbf{k},+}),
$
where 
$
m_\pm = 2m_\uparrow m_\downarrow/(m_\downarrow \pm m_\uparrow),
$
$V$ is the volume of the system, and
$
\epsilon_{\mathbf{k},\pm} = (\epsilon_{\mathbf{k},\uparrow} \pm \epsilon_{\mathbf{k},\downarrow})/2.
$
Note that $m_+$ is twice the reduced mass of the $\uparrow$ and 
$\downarrow$ fermions, and that the equal mass case corresponds to 
$|m_-| \to \infty$. Here, and throughout, the momentum space sums 
are evaluated as
$
\sum_\mathbf{k} \equiv [V/(2\pi)^3] \int d^3\mathbf{k} \equiv [V/(2\pi^2)] \int_0^{\pi/d_z} dk_z \int_0^\infty k_\perp dk_\perp,
$
since the system at hand has a cylindrical symmetry around the $k_z$-axis, 
and a translational symmetry along the $\mathbf{\widehat{z}}$ direction, 
so that the $k_z$ integrals are limited to the first Brillouin zone, 
i.e. $-\pi/d_z \le k_z \le \pi/d_z$. The resultant integrands are also 
even functions of $k_z$, and hence we integrate over half of the Brillouin 
zone and multiply them by two.

In this manuscript, for its simplicity, we set $t_\downarrow = 1/(2m_\downarrow d_z^2)$ 
but allow for the $\uparrow$ fermions to have a different effective mass along the 
lattice direction through a tight-binding dispersion determined by $t_\uparrow$,
so that
\begin{eqnarray}
\epsilon_{\mathbf{k},\uparrow} &=& \frac{k_\perp^2}{2m_\uparrow} + 2t_\uparrow \left[1 - \cos(k_z d_z)\right],
\label{eqn:epsup} \\
\epsilon_{\mathbf{k},\downarrow} &=& \frac{k_\perp^2}{2m_\downarrow} + \frac{1}{m_\downarrow d_z^2} \left[1 - \cos(k_z d_z)\right].
\label{eqn:epsdo} 
\end{eqnarray}
Notice that when $k_{Fz, \downarrow} d_z \ll \pi$ (i.e. the low-filling
limit for the $\downarrow$ species), where $k_{Fz, \downarrow}$ is the 
Fermi momentum of $\downarrow$ fermions in the $k_z$ direction, 
since Eq.~(\ref{eqn:epsdo}) can be approximated as 
$\epsilon_{\mathbf{k},\downarrow} \approx k^2/(2m_\downarrow)$,
optical lattice has negligibe effect on $\downarrow$ fermions in this limit.
Next, we analyze the phase diagram of the Hamiltonian given 
in Eq.~(\ref{eqn:hamiltonian}) with the dispersions given by Eqs.~(\ref{eqn:epsup}) 
and~(\ref{eqn:epsdo}), within the saddle-point (mean-field) approximation.

\subsection{Self-consistency equations}
\label{sec:self-consistency}
At low temperatures ($T \approx 0$), the saddle-point self-consistency 
(order parameter and number) equations are sufficient to describe 
the BCS-BEC evolution of superfluidity~\cite{leggett, jan}. 
For the Hamiltonian given in Eq.~(\ref{eqn:hamiltonian}), the 
saddle-point action is $S_0 = \Omega_0/T$, where
\begin{eqnarray}
\Omega_0 = \frac{|\Delta|^2}{g} 
+ \sum_\mathbf{k} \left( \xi_{\mathbf{k},+} - E_{\mathbf{k},+} \right)
+ T \sum_{\mathbf{k},s} \ln [f(-E_{\mathbf{k},s})] 
\end{eqnarray}
is the saddle-point thermodynamic potential. 
Here,
$
f(x) = 1/[\exp(x/T) + 1]
$
is the Fermi function,
$
E_{\mathbf{k},s} = (\xi_{\mathbf{k},+}^2 + |\Delta|^2)^{1/2} + \gamma_s \xi_{\mathbf{k},-}
$
is the quasiparticle energy when $\gamma_1 = 1$ or
the negative of the quasihole energy when $\gamma_2 = -1$, and
$
E_{\mathbf{k},\pm} = (E_{\mathbf{k},1} \pm E_{\mathbf{k},2})/2.
$
In addition, $\Delta$ is the order parameter and
$
\xi_{\mathbf{k},\pm} = \epsilon_{\mathbf{k},\pm} - \mu_\pm,
$
where 
$
\mu_\pm = (\mu_\uparrow \pm \mu_\downarrow)/2.
$
Note that the symmetry between quasiparticles and quasiholes 
is broken when $\xi_{\mathbf{k},-} \ne 0$.

The saddle-point condition $\delta \Omega_0 /\delta \Delta^* = 0$ leads 
to an equation for the order parameter,
\begin{equation}
- \frac{m_+ V}{4\pi a_{eff}} = \sum_{\mathbf{k}} \left[ \frac{1 - f(E_{\mathbf{k},1}) - f(E_{\mathbf{k},2})}
{2E_{\mathbf{k},+}} - \frac{1}{2\epsilon_{\mathbf{k},+}} \right],
\label{eqn:op}
\end{equation}
where, as usual, $g$ is eliminated in favor of the effective $s$-wave 
scattering length $a_{eff}$ via the relation given above~\cite{leggett, jan}.
The order parameter equation has to be solved self-consistently 
with the number equations. At the saddle point, the relation
$N_\sigma = -\partial \Omega_0/\partial {\mu_\sigma}$ leads to
\begin{eqnarray}
N_{\uparrow} &=& \sum_{\mathbf{k}} \left[ |u_{\mathbf{k}}|^2 f(E_{\mathbf{k},1})+ |v_{\mathbf{k}}|^2 f(-E_{\mathbf{k},2}) \right],
\label{eqn:nup} \\
N_{\downarrow} &=& \sum_{\mathbf{k}} \left[ |u_{\mathbf{k}}|^2 f(E_{\mathbf{k},2})+ |v_{\mathbf{k}}|^2 f(-E_{\mathbf{k},1}) \right],
\label{eqn:ndo}
\end{eqnarray}
where
$
|u_{\mathbf{k}}|^2 = (1 + \xi_{\mathbf{k}, +}/E_{\mathbf{k}, +})/2
$
and
$
|v_{\mathbf{k}}|^2 = (1 - \xi_{\mathbf{k}, +}/E_{\mathbf{k}, +})/2
$
are the usual coherence factors. At $T = 0$, Eqs.~(\ref{eqn:op}),~(\ref{eqn:nup})
and~(\ref{eqn:ndo}) can be simplified considerably as shown in
Appendix A.

In order to analyze the phase diagram at $T = 0$, we solve the 
saddle-point self-consistency equations and check the stability of 
these solutions for the uniform superfluid phase using the 
compressibility (or the curvature) criterion~\cite{iskin06, he06, pao06}. 
This says that the compressibility matrix $\mathbf{\kappa}(T)$ with elements
$
\kappa_{\sigma,\sigma'} (T) = - \partial^2 \Omega_0 / (\partial \mu_\sigma \partial \mu_{\sigma'})
$
needs to be positive definite, and it is related (identical) to the 
condition that the curvature
\begin{equation}
\frac{\partial^2 \Omega_0} {\partial \Delta^2} = 
|\Delta|^2 \sum_{\mathbf{k},s} 
\left[
\frac{0.5-f(E_{\mathbf{k},s})}{E_{\mathbf{k},+}^3}
+ \frac{f'(E_{\mathbf{k},s})}{E_{\mathbf{k},+}^2}  
\right]
\label{eqn:curvature}
\end{equation}
of the saddle-point thermodynamic potential $\Omega_0$ with respect 
to the saddle-point parameter $\Delta$ needs to be positive. 
Here, $f'(x) = df(x)/dx$.
When at least one of the eigenvalues of $\mathbf{\kappa} (T)$, 
or the curvature $\partial^2 \Omega_0 /\partial \Delta^2$ is
negative, the uniform saddle-point solution does not correspond to 
a minimum of $\Omega_0$, and a nonuniform superfluid phase, e.g. 
a phase separation, is favored.

\subsection{Noninteracting Limit}
\label{sec:nil}

Before presenting our numerical results, let's first analyze the
noninteracting $g \to 0$ limit. In this limit, since $1/a_{eff} \to -\infty$,
the order parameter vanishes $\Delta \to 0$, and at $T = 0$ 
Eqs.~(\ref{eqn:nup}) and~(\ref{eqn:ndo}) can be written as
$
N_{\sigma} = \sum_{\mathbf{k}} \theta(-\xi_{\mathbf{k},\sigma}),
$
where $\theta(x)$ is the heaviside step function. Evaluating the 
$\mathbf{k}$-space sums, the density $n_\sigma = N_\sigma/V$ 
of the $\sigma$ fermions become
\begin{eqnarray}
n_\sigma &=& \frac{m_\sigma k_{Fz,\sigma}}{2\pi^2} (\mu_\sigma - 2t_\sigma)  
+ \frac{m_\sigma t_\sigma}{\pi^2 d_z} \sin(k_{Fz,\sigma} d_z), 
\label{eqn:nsigmani} \\
\mu_\sigma &=& 2t_\sigma \left[ 1 - \cos(k_{Fz,\sigma} d_z) \right],
\label{eqn:musigmani}
\end{eqnarray}
where $k_{Fz,\sigma} \geq 0$ corresponds to the Fermi momentum of the $\sigma$ 
fermions in the $k_z$ direction, which is related to the chemical 
potential via Eq.~(\ref{eqn:musigmani}).

Note that, in the low-density ($k_{Fz,\downarrow} d_z \ll \pi$) limit, expanding out
Eqs.~(\ref{eqn:nsigmani}) and~(\ref{eqn:musigmani}) to the lowest nontrivial
orders (third and second orders, respectively) in $k_{Fz,\downarrow} d_z$, we obtain
$
n_\downarrow \approx m_\downarrow k_{Fz,\downarrow} (\mu_\downarrow - t_\downarrow k_{Fz,\downarrow}^3 d_z^2/3)/ (2\pi^2)
$
and $\mu_\downarrow \approx t_\downarrow k_{Fz,\downarrow}^2 d_z^2$. Combining these 
two expressions, and setting $k_{Fz,\downarrow} = k_{F,\downarrow}$ 
or $t_\downarrow = 1/(2m_\downarrow d_z^2)$, the density of $\downarrow$ 
fermions acquires the usual form
$
n_\downarrow \approx k_{F,\downarrow}^3/(6\pi^2).
$
In addition, when $t_\uparrow \to 0$ and $k_{Fz,\uparrow} = \pi/d_z$, 
Eq.~(\ref{eqn:nsigmani}) reduces to the density of $\uparrow$ fermions 
in each (isolated) two-dimensional planes along the $k_z$ direction,
$
N_\uparrow/(M A) =  m_\uparrow k_{F,\uparrow}^2/(4\pi),
$
which is of the usual form, where we used $V = A L_z$ with $A$ being 
the area of the system in the $(x,y)$ plane and $L_z$ is the system 
size in the $\mathbf{\widehat{z}}$ direction.
Here, $M = L_z/d_z$ is the number of two-dimensional planes, i.e. the
number of lattice sites, in the $\mathbf{\widehat{z}}$ direction. 
Having discussed the noninteracting limit, next we analyze the 
BCS-BEC evolution.

\section{Saddle-point Approximation}
\label{sec:sps}

In this section, we consider only equal density mixtures, where 
$n_\uparrow = n_\downarrow$, at zero temperature. For these mixtures, 
first we analyze the amplitude of the order parameter $|\Delta|$, 
chemical potential sum $\mu_+$, and the chemical potential 
difference $\mu_-$ as a function of the tunneling amplitude 
$t_\uparrow$ and effective scattering length $1/(k_{F,\downarrow} a_{eff})$
for fixed values of $k_{F,\downarrow} d_z$. Here, $k_{F,\downarrow}$ 
is an ``effective'' Fermi momentum for $\downarrow$ fermions defined through the 
three-dimensional density $n_\downarrow = k_{F,\downarrow}^3/(6\pi^2)$,
where $n_\downarrow$ is given by Eq.~(\ref{eqn:nsigmani}).
Then, using the stability criterion given in Eq.~(\ref{eqn:curvature}), we 
construct the phase diagrams. For its simplicity, we mainly present our results 
for the equal mass ($m_\uparrow = m_\downarrow$) mixtures, but we also 
briefly mention the effects of mass anisotropy $m_\uparrow \ne m_\downarrow$ 
on the phase diagrams.

\begin{figure} [htb]
\centerline{\scalebox{0.6}{\includegraphics{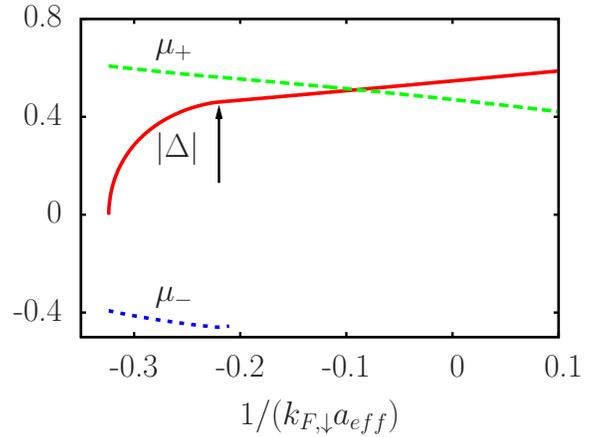}}}
\caption{\label{fig:gap} (Color online)
Saddle-point solutions for the amplitude of the order parameter 
$|\Delta|$, chemical potential sum $\mu_+ = (\mu_\uparrow + \mu_\downarrow)/2$, 
and the chemical potential difference $\mu_- = (\mu_\uparrow - \mu_\downarrow)/2$ 
are shown (in units of $\epsilon_{F,\downarrow}$) as a function of 
the effective $s$-wave scattering parameter $1/(k_{F,\downarrow} a_{eff})$.
($\mu_-$ is shown only for weak interactions where it is relevant). 
This data corresponds to the case where $m_\uparrow = m_\downarrow$, 
$t_\uparrow = \epsilon_{F,\downarrow}$ and $k_{F,\downarrow} d_z = 0.1$.
The arrow shows the location of the topological gapless superfluid 
to gapped superfluid transition discussed in the text.
}
\end{figure}
\subsection{Normal-Superfluid transition}
\label{sec:nst}

Using $k_{F,\downarrow}$ and $\epsilon_{F,\downarrow} = k_{F,\downarrow}^2/(2m_\downarrow)$ 
as our length and energy scales, respectively, we solve Eqs.~(\ref{eqn:op0}),~(\ref{eqn:nup0})
and~(\ref{eqn:ndo0}) numerically. For instance, in Fig.~\ref{fig:gap}, we show 
self-consistent solutions of $|\Delta|$, $\mu_+$ and $\mu_-$ as a 
function of $1/(k_{F,\downarrow} a_{eff})$, when $m_\uparrow = m_\downarrow$, 
$t_\uparrow = \epsilon_{F,\downarrow}$ and $k_{F,\downarrow} d_z = 0.1$.
When the scattering parameter is smaller than 
a critical value, i.e. $1/(k_{F,\downarrow} a_{eff}) \lesssim -0.32$, 
the saddle-point solution $\mu_\downarrow = \mu_+ - \mu_- \to 1$ 
(in units of $\epsilon_{F,\downarrow}$) since $|\Delta| = 0$, indicating 
that the mixture is a normal Fermi gas~\cite{nishidanote}. Beyond this critical value, 
the superfluid order parameter $|\Delta|$ becomes nonzero indicating a 
quantum phase transition from the normal to a superfluid phase. 

This transition can be understood from earlier works on Cooper pairing 
with mismatched Fermi surfaces. For instance, in the case of Fermi gases
with unequal densities in purely three-dimensions, a superfluid to normal
phase transition has been recently observed beyond a critical density 
difference depending on the value of the scattering 
parameter~\cite{zwierlerin06, partridge06, shin08, salomon10}. This transition 
occurs when the difference in the chemical potentials (or Fermi momenta) 
reaches what is known as the Clogston-Chandrasekhar limit~\cite{clogston, chandrasekhar}. 
In our case, the main mechanism is the same. As can be extracted from 
Eq.~(\ref{eqn:musigmani}), the mismatch is 
inevitable in some parts of the $\mathbf{k}$-space even for the equal 
density mixtures considered in this manuscript. Therefore, it is energetically 
more favorable for the mixture to be in the normal phase until a 
critical scattering parameter is reached, beyond which Cooper pairing is possible.

\begin{figure} [htb]
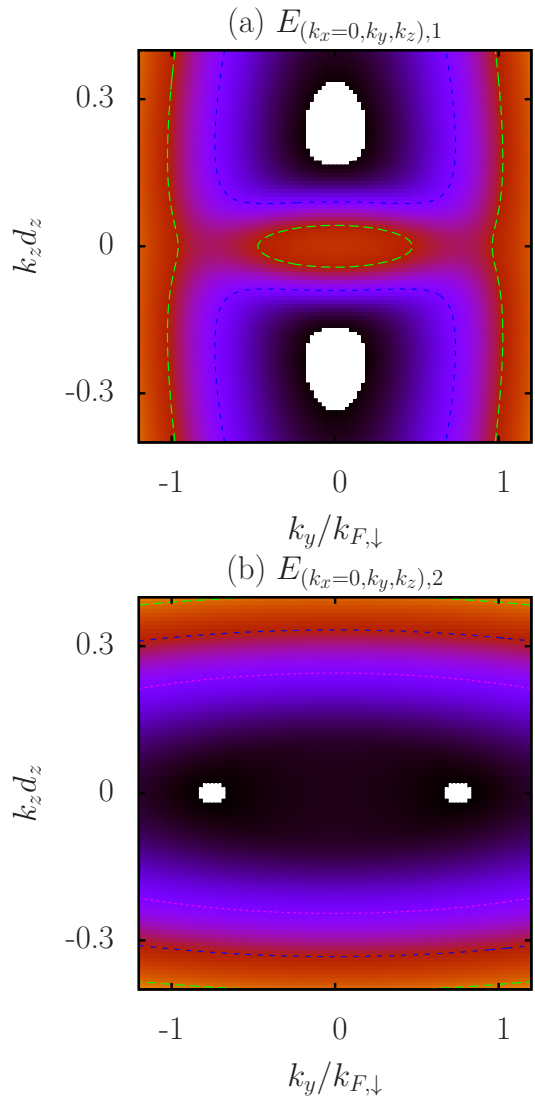

\centerline{\scalebox{0.6}{\includegraphics{fig2a.epsi}}}
\centerline{\scalebox{0.6}{\includegraphics{fig2b.epsi}}}
\caption{\label{fig:ek} (Color online)
Contour maps of (a) the quasiparticle excitation energy $E_{\mathbf{k},1}$, and 
(b) the negative of the quasihole excitation energy $E_{\mathbf{k},2}$ are 
shown as a function of momentum components $k_z$ (in units of $1/d_z$) and 
$k_y$ (in units of $k_{F,\downarrow}$) in the $k_x = 0$ plane. 
This data corresponds to the case where $m_\uparrow = m_\downarrow$, 
$t_\uparrow = \epsilon_{F,\downarrow}$, $k_{F,\downarrow} d_z = 0.1$ and
$1/(k_{F,\downarrow} a_{eff}) = -0.25$.
The excitations are gapless ($E_{\mathbf{k},s} \le 0$) in the white regions.
}
\end{figure}
\subsection{Topological gapless-gapped superfluidity transition}
\label{sec:topological}

With further increase in the scattering parameter, 
$|\Delta|$ increases quite rapidly and $\mu_+$ decreases, with a
kink in the former quantity at $1/(k_{F,\downarrow} a_{eff}) \approx -0.22$. 
(We also expect a weak kink in $\mu_+$ at the same point, but it is not
clearly seen in the data.) Therefore, the BCS-BEC evolution in 
mixed-dimensional Fermi gases is nonanalytic, i.e. it is not a crossover. 
Recall that, in usual three-dimensional mixtures, the evolution of 
$|\Delta|$ and $\mu_+$ is analytic for all $1/(k_{F,\downarrow} a_{eff})$, 
and the evolution is just a crossover. The kink in $|\Delta|$ is more pronounced 
for lower values of $t_\uparrow$, and it signals a topological quantum 
phase transition as discussed next. 

The excitation spectrum of quasiparticles are determined by energies 
$E_{\mathbf{k},1}$ and $E_{\mathbf{k},2}$. At $\mathbf{k}$-space points, 
the condition $E_{\mathbf{k},s} = 0$ defines Fermi surfaces of quasiparticles 
in momentum space where the quasiparticle excitation spectrum changes 
from a gapped to a gapless phase. These changes in the Fermi surfaces 
of quasiparticles are topological in nature~\cite{volovik}, and we identify 
topological quantum phase transitions associated with the disappearance 
or appearance of momentum space regions of zero quasiparticle energies 
when either $1/(k_{F,\downarrow} a_{eff})$, $t_\uparrow$,
and/or $k_{F,\downarrow} d_z$ is changed. Note that the topological 
transition occurs without changing the symmetry of the order parameter 
as the Landau classification demands for ordinary phase transitions. 

We illustrate the gapless superfluid phase in Fig.~\ref{fig:ek}, where contour 
maps of $E_{\mathbf{k},s}$ are shown as a function of $k_z$ and $k_y$ in the $k_x = 0$ plane, 
when $m_\uparrow = m_\downarrow$, $t_\uparrow = \epsilon_{F,\downarrow}$, 
$k_{F,\downarrow} d_z = 0.1$ and $1/(k_{F,\downarrow} a_{eff}) = -0.25$.
In Fig.~\ref{fig:gap}, this data corresponds to a point that is slightly 
on the left hand side of the transition point indicated by an arrow. 
The excitations are gapless ($E_{\mathbf{k},s} \le 0$) in the white regions.

\begin{figure} [htb]
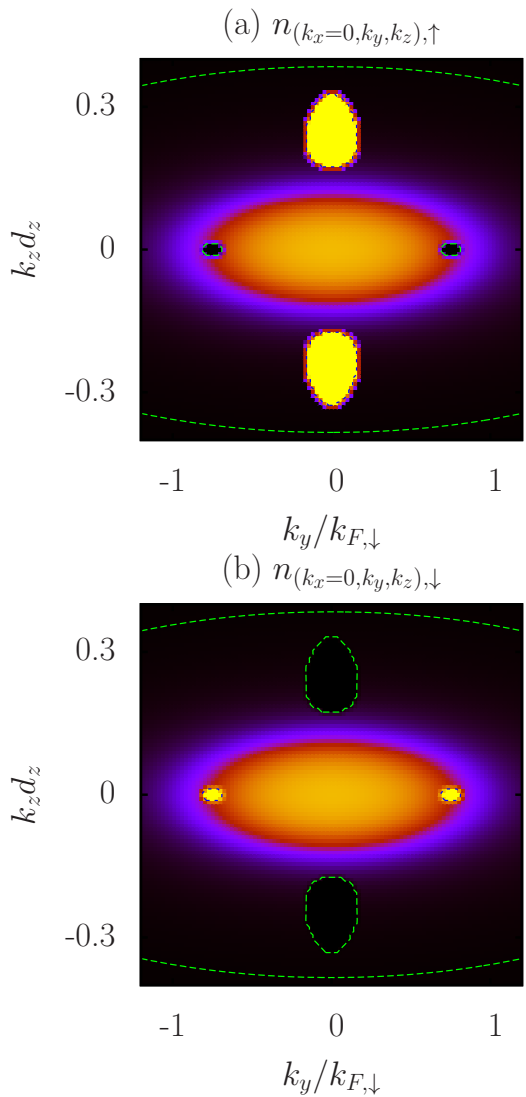

\centerline{\scalebox{0.62}{\includegraphics{fig3a.epsi}}}
\centerline{\scalebox{0.62}{\includegraphics{fig3b.epsi}}}
\caption{\label{fig:nk} (Color online)
Contour maps of the momentum distributions 
(a) $n_{\mathbf{k},\uparrow}$ for the $\uparrow$ fermions, 
and (b) $n_{\mathbf{k},\downarrow}$ for the $\downarrow$ fermions are 
shown as a function of momentum components $k_z$ (in units of $1/d_z$) and 
$k_y$ (in units of $k_{F,\downarrow}$) in the $k_x = 0$ plane.
This data corresponds to the case considered in Fig.~\ref{fig:ek}, 
where $m_\uparrow = m_\downarrow$, $t_\uparrow = \epsilon_{F,\downarrow}$, 
$k_{F,\downarrow} d_z = 0.1$ and $1/(k_{F,\downarrow} a_{eff}) = -0.25$.
Note that when $E_{{\bf k},1} > 0$ and $E_{\mathbf{k},2} > 0$,
then $n_{\mathbf{k},\uparrow} =  n_{\mathbf{k},\downarrow}$.
However, when $E_{\mathbf{k},1} \le 0$ and $E_{\mathbf{k},2} > 0$, then
$n_{\mathbf{k},\uparrow} = 1$ and $n_{\mathbf{k},\downarrow} = 0$;
and when $E_{\mathbf{k},1} > 0$ and $E_{\mathbf{k},2} \le 0$, then
$n_{\mathbf{k},\uparrow} = 0$ and $n_{\mathbf{k},\downarrow} = 1$. 
The densities are 1 (0) in the bright yellow (black) regions.
}
\end{figure}

The topological transition could be potentially observed through the 
measurement of the momentum distribution $n_{\mathbf{k},\sigma}$ of the 
fermions~\cite{md}, which can be extracted from Eqs.~(\ref{eqn:nup}) and~(\ref{eqn:ndo}). 
For $\mathbf{k}$-space regions where $E_{{\bf k},1} > 0$ and 
$E_{\mathbf{k},2} > 0$, the corresponding momentum distributions are equal 
$n_{\mathbf{k},\uparrow} =  n_{\mathbf{k},\downarrow}$.
However, when $E_{\mathbf{k},1} \le 0$ and $E_{\mathbf{k},2} > 0$, then
$n_{\mathbf{k},\uparrow} = 1$ and $n_{\mathbf{k},\downarrow} = 0$. 
Similarly, when $E_{\mathbf{k},1} > 0$ and $E_{\mathbf{k},2} \le 0$, then
$n_{\mathbf{k},\uparrow} = 0$ and $n_{\mathbf{k},\downarrow} = 1$. 
We illustrate these cases in Fig.~\ref{fig:nk} for the parameters of 
Fig.~\ref{fig:ek}. Note that, although there are excess (or unpaired) 
$\uparrow$ or $\downarrow$ fermions in different regions of the 
$\mathbf{k}$-space, i.e. the bright yellow regions, there are equal 
number of $\uparrow$ and $\downarrow$ fermions in total. The size of yellow 
regions may look very different in (a) and (b) due partly to the 
difference in scaling factors in $k_z$ and $k_y$. 

This topological transition is quantum ($T=0$) in nature, but its
signatures should still be observed at finite temperatures within the 
quantum critical region, where the momentum distributions are smeared 
out due to thermal effects. In addition, while thermodynamic quantities 
such as atomic compressibility, specific heat, and spin susceptibility 
have power-law dependences on the temperature in the BCS side,
they have exponential dependences on the temperature and
the minimum energy of quasiparticle excitations in the BEC side, again
signaling the existence of a quantum phase transition at $T = 0$.
Having discussed the topological classification of possible superfluid 
phases, we are ready to present the saddle-point phase diagrams 
at $T = 0$, including the stability analysis (positive curvature 
criterion) of the solutions.

\subsection{Ground state phase diagrams}
\label{sec:gspd}

\begin{figure} [htb]
\centerline{\scalebox{0.6}{\includegraphics{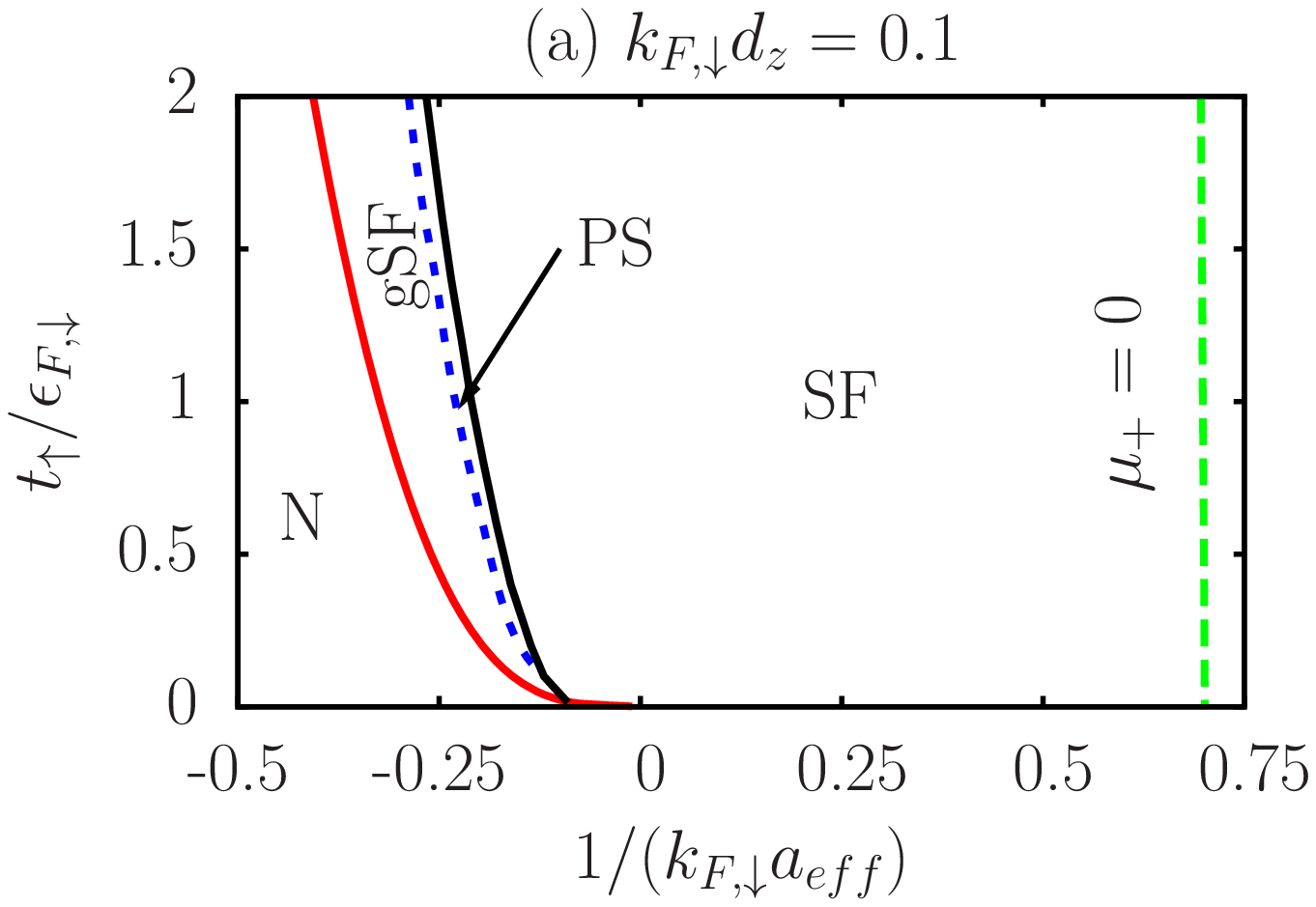}}}
\centerline{\scalebox{0.6}{\includegraphics{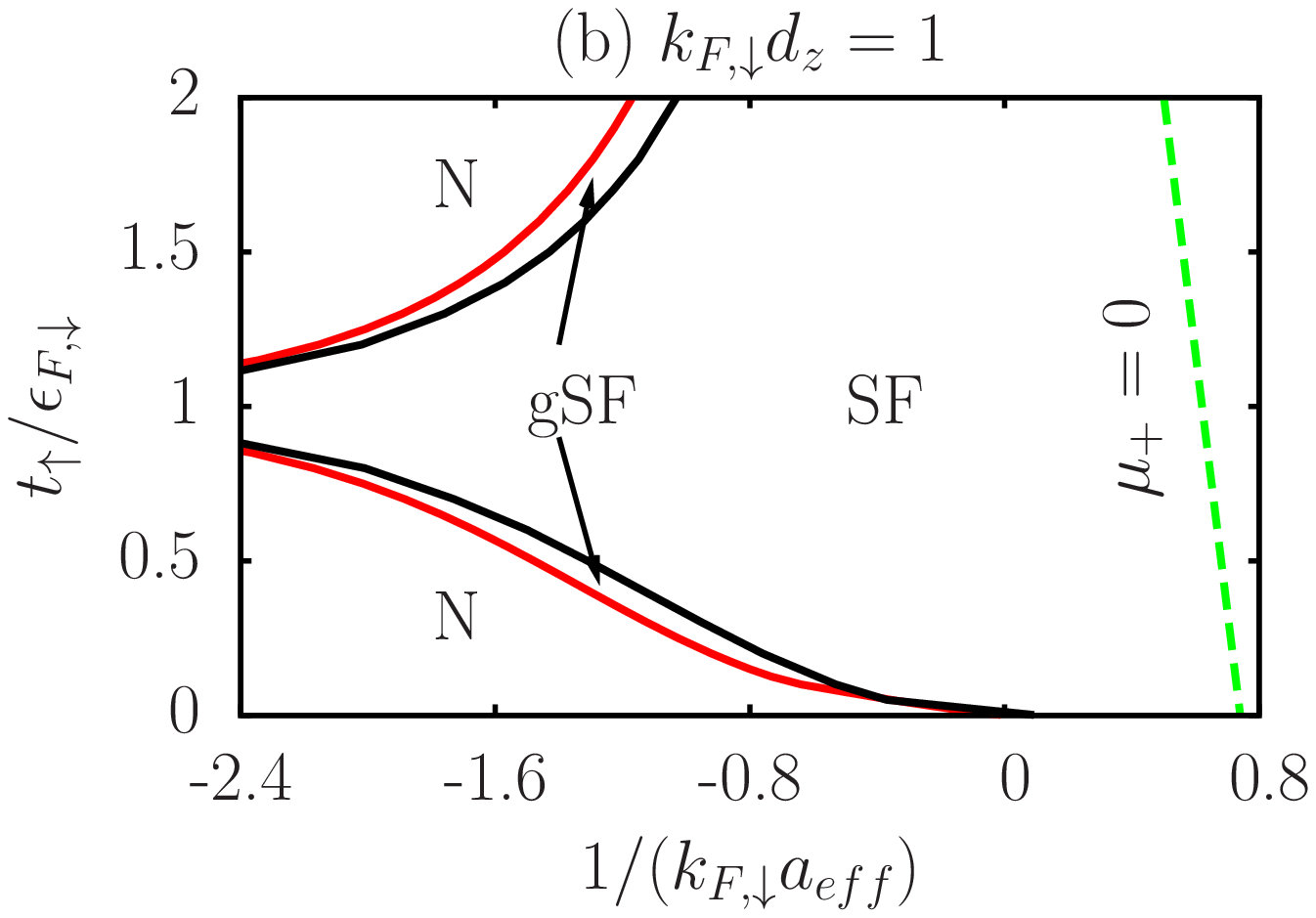}}}
\caption{\label{fig:pd} (Color online)
Ground state phase diagrams are shown as a function of the tunneling 
amplitude $t_\uparrow$ (in units of $\epsilon_{F,\downarrow}$) 
and the effective $s$-wave scattering parameter $1/(k_{F,\downarrow} a_{eff})$.
Here, we consider equal density ($n_\uparrow = n_\downarrow$) 
and equal mass ($m_\uparrow = m_\downarrow$) mixtures, for fixed values of
$k_{F,\downarrow} d_z = 0.1$ in (a), and $k_{F,\downarrow} d_z = 1$ in (b).
We show normal ($N$), gapped uniform superfluid ($SF$), 
gapless uniform superfluid ($gSF$), 
and phase-separated ($PS$) regions. The $PS$ region in (b)
is close to the normal-superfluid transition boundary, but it is very 
small and not shown. There is no phase transition across $\mu_+ = 0$ 
(green-dashed) line, it is only shown as a reference point.
}
\end{figure}

In Fig.~\ref{fig:pd}, ground state phase diagrams are shown as a 
function of the tunneling amplitude $t_\uparrow$ and the effective 
$s$-wave scattering parameter $1/(k_{F,\downarrow} a_{eff})$ for fixed
values of (a) $k_{F,\downarrow} d_z = 0.1$, and (b) $k_{F,\downarrow} d_z = 1$. 
We indicate normal ($N$), gapped uniform superfluid ($SF$), 
gapless uniform superfluid ($gSF$), and phase-separated ($PS$) regions. 
The normal phase is characterized by a vanishing order parameter
($\Delta = 0$), while the gapped superfluid and gapless superfluid 
phases are both characterized by $\Delta > 0$ 
and $\partial^2 \Omega_0 /\partial \Delta^2 > 0$,
but with distinct $\mathbf{k}$-space topologies as discussed above.
In the $gSF$ phase, the unpaired $\uparrow$ and $\downarrow$ fermions 
coexist with the paired (or superfluid) ones in different 
$\mathbf{k}$-space regions, but there are no unpaired fermions 
in the $SF$ phase, i.e. all $\uparrow$ and $\downarrow$ fermions are paired.
The $gSF$ phase is in some ways similar to the Sarma state 
found in mixtures with unequal densities~\cite{sarma}, but in our case, the 
gapless superfluid phase is unpolarized and most importantly it is 
stable against phase separation. The phase-separated region is 
characterized by $\partial^2 \Omega_0 /\partial \Delta^2 < 0$, 
but this region could also be of the FFLO-type superfluid having 
spatial modulations~\cite{FF,LO}. Such a possibility is not 
considered in this manuscript, and it is left as an important 
problem to address in the future.

We can understand these phase diagrams as follows. 
For a fixed $t_\uparrow$, when the scattering parameter is smaller 
than a critical value, the potential energy is not sufficient 
to cause pairing due to mismatch of the Fermi surfaces, 
and the mixture is a normal Fermi gas with $\Delta = 0$. 
As shown in Fig.~\ref{fig:pd}(a), the critical scattering parameter 
decreases with increasing $t_\uparrow$, since increasing $t_\uparrow$ 
decreases the mismatch for lower values of $t_\uparrow$. 
In the normal region, when $t_\uparrow = 0$, Fermi surface of 
the $\uparrow$ fermions is a cylindrical shell in the $\mathbf{k}$-space 
with height $k_{Fz,\uparrow} = 2\pi/d_z$ and radius $k_{F,\uparrow}$. 
However, in the low-filling limit of $\downarrow$ fermions, 
Fermi surface of the $\downarrow$ fermions is more like a spherical 
shell with radius $k_{F,\downarrow}$. 
Note that for equal mass and equal density mixtures considered here, 
the $\mathbf{k}$-space volumes enclosed by the cylindrical and 
spherical Fermi surfaces must be equal. Therefore, at $t_\uparrow = 0$, 
there is a large mismatch between the two Fermi surfaces when
$k_{F,\uparrow} \ll k_{F,\downarrow} \ll \pi/d_z$, and increasing 
$t_\uparrow$ from zero decreases $k_{Fz,\uparrow}$ and increases the ratio 
$k_{F,\uparrow}/k_{F,\downarrow}$. When this happens, 
Fermi surface of the $\uparrow$ fermions looks like a prolate 
spheroid (like an american football). This decreases the mismatch 
for small values of $t_\uparrow$ as long as $k_{F,\uparrow} \lesssim k_{F,\downarrow}$, 
and it is qualitatively what happens along the normal-superfluid 
transition boundary in Fig.~\ref{fig:pd}(a), 
when $k_{F,\downarrow} d_z = 0.1 \ll \pi$. 

In Fig.~\ref{fig:pd}(b), we show the same phase diagram for a higher value
of $k_{F,\downarrow} d_z = 1$. Similar to Fig.~\ref{fig:pd}(a), 
the critical scattering parameter for the normal-superfluid transition 
decreases initially with increasing $t_\uparrow$, since increasing $t_\uparrow$ 
decreases the mismatch for lower values of $t_\uparrow$. 
In contrast, beyond a critical value of $t_\uparrow$ 
[e.g. $t_\uparrow \gtrsim \epsilon_{F,\downarrow}$ in Fig.~\ref{fig:pd}(b)],
the critical scattering parameter increases with increasing $t_\uparrow$.
This is because, since increasing $t_\uparrow$ decreases $k_{F,z}$ and 
increases the ratio $k_{F,\uparrow}/k_{F,\downarrow}$, it eventually
increases the mismatch of the two Fermi surfaces after
$k_{F,\uparrow} > k_{F,\downarrow}$. When this happens, the 
Fermi surface of the $\uparrow$ fermions changes from a prolate spheroid 
to an oblate spheroid (a disk-shaped ellipsoid). For the case 
considered in Fig.~\ref{fig:pd}(a) where $k_{F,\downarrow} d_z = 0.1$, 
we found that this occurs beyond $t_\uparrow \gtrsim 100 \epsilon_{F,\downarrow}$, 
but it is not shown.
In general, for equal mass and equal density mixtures, such a change 
is expected to occur beyond $t_\uparrow \gtrsim 1/(2 m_\uparrow d_z^2)$,
and this expectation is consistent with our numerical findings.

In addition, in Fig.~\ref{fig:pd}, the solid-black lines correspond to
the transition boundary between the gapless superfluid ($gSF$) and gapped 
superfluid ($SF$) phases. However, the stability criterion 
given in Eq.~(\ref{eqn:curvature}) is not satisfied in some parts 
of the gapless superfluid region, indicating a phase separation.
This occurs between the dashed-blue and the solid-black lines in Fig.~\ref{fig:pd}(a). 
In contrast, phase separation occurs in a tiny region very close to 
the normal-superfluid transition boundary (solid-red line) 
in Fig.~\ref{fig:pd}(b), and it is not shown.
The gapless superfluid phase phase is in some ways similar to 
the Sarma state found in mixtures with unequal densities~\cite{sarma}, 
but in our case, $gSF$ is unpolarized and most importantly it is 
stable against phase separation in a considerably large region as
shown in the figures. In this way, our $gSF$ phase is similar to
those of Refs.~\cite{eite, feigun}, which are recently proposed  
for ultracold atomic systems in other contexts. 

Before concluding, we would like to comment on the phase diagram 
of mixed-dimensional two-species Fermi-Fermi mixtures.
When $\uparrow$ and $\downarrow$ fermions have different masses, 
the phase boundaries shift left (right) when the $\uparrow$ species 
is heavier (lighter) than the $\downarrow$ fermions. 
For instance, in the case of $^6$Li-$^{40}$K mixtures, 
the $t_\uparrow \to 0$ limit of the normal-superfluid 
boundary for the case considered in Fig.~\ref{fig:pd}(a) shifts 
to $1/(k_{F,\downarrow} a_{eff}) \approx -0.6$ when 
$m_\uparrow = 6.64 m_\downarrow$ (when $^{40}$K atoms correspond to $\uparrow$) 
and to $1/(k_{F,\downarrow} a_{eff}) \approx 0.6$ when 
$m_\uparrow = 0.15 m_\downarrow$ (when $^6$Li-atoms correspond to $\uparrow$).

\section{Conclusions}
\label{sec:conclusions}

Motivated by a very recent experiment involving 
mixed-dimensional Bose-Bose mixtures~\cite{lamporesi}, here we 
investigated the ground state phase diagram of superfluidity for 
mixed-dimensional Fermi-Fermi mixtues in the BCS-BEC evolution. 
In this recent experiment, a species-selective one-dimensional 
optical lattice is applied to a two-species mixture of bosonic atoms,
such that only one of the species feel the lattice potential,
and is confined to a quasi-two-dimensional geometry, 
while having negligible effect on the other, that is leaving it 
three dimensional. We considered a similar problem with 
two-species mixtures of fermionic atoms, where both species are 
confined to quasi-two-dimensional geometries determined by their 
hoppings along the lattice direction. 

We considered equal-density mixtures at zero temperature, and after
solving the saddle-point self-consistency equations, we constructed
the phase diagrams using some stability criterion. We found normal, 
gapped superfluid, gapless superfluid, and phase separated regions.
The gapped superfluid and gapless superfluid phases are 
identified with the disappearance or appearance of momentum 
space regions of zero quasiparticle energies.
In particular, we found a stable gapless superfluid phase 
where the unpaired $\uparrow$ and $\downarrow$ fermions coexist 
with the paired (or superfluid) ones in different 
momentum space regions. This phase is in some ways similar 
to the Sarma state found in mixtures with unequal densities~\cite{sarma}, 
but in our case, the gapless superfluid phase is unpolarized and 
most importantly it is stable against phase separation. 
We also argued that the topological transition from the gapped 
superfluid to the gapless superfluid could be potentially observed 
through the measurement of the momentum distribution~\cite{md}.

There are several ways to extend this work. First, the possibility 
of FFLO-type superfluid phases~\cite{FF,LO}, where Cooper pairs 
have finite center of mass momentum, leading to a spatially 
modulated phase, is not considered in this manuscript. This is 
an important problem to address due to its relevance to 
condensed-matter systems. 
Second, our calculation is based on the saddle-point self-consistency 
equations, which are known to be sufficient to qualitatively 
describe the entire BCS-BEC evolution, at least for the usual 
three-dimensional mixtures at low temperatures. 
However, corrections beyond the saddle point could be important 
in stabilizing or destablizing especially the gapless superfluid 
phase~\cite{saddle}. Third, we used a single-band model to describe 
the optical lattice potential, and the effects of higher bands 
could become important near the strongly interacting regime. 
Lastly, atomic systems are not uniform since confining trapping 
potentials are always present, and finite-size 
effects due to such potentials could also be analyzed.

\section{Acknowledgments}
\label{sec:ack}

We thank the Scientific and Technological Research Council of Turkey 
(T\"{U}B$\dot{\mathrm{I}}$TAK) for financial support, and 
Institute of Theoretical and Applied Physics (ITAP - Marmaris) 
for their hospitality.

\appendix

\section{Self-consistency equations at $T = 0$}
\label{sec:app.a}

For numerical purposes, the self-consistency equations can be simplified
as follows. At zero temperature, since the Fermi function $f(x)$ turns
into a heaviside step function $\theta(-x)$, Eqs.~(\ref{eqn:op}),
~(\ref{eqn:nup}) and~(\ref{eqn:ndo}) become
\begin{eqnarray}
\frac{m_+ V}{4\pi a_{eff}} &=& \sum_{\mathbf{k}} 
\left[ \frac{1}{2\epsilon_{\mathbf{k},+}} 
- \frac{1 - \theta(-E_{\mathbf{k},1}) - \theta(-E_{\mathbf{k},2})}
{2E_{\mathbf{k},+}} \right],
\label{eqn:op0} \\
N_{\uparrow} &=& \sum_{\mathbf{k}} \left[ |u_{\mathbf{k}}|^2 \theta(-E_{\mathbf{k},1})+ |v_{\mathbf{k}}|^2 \theta(E_{\mathbf{k},2}) \right],
\label{eqn:nup0} \\
N_{\downarrow} &=& \sum_{\mathbf{k}} \left[ |u_{\mathbf{k}}|^2 \theta(-E_{\mathbf{k},2})+ |v_{\mathbf{k}}|^2 \theta(E_{\mathbf{k},1}) \right],
\label{eqn:ndo0}
\end{eqnarray}
where the $\mathbf{k}$-space sums are,
$
\sum_\mathbf{k} \equiv [V/(2\pi)^3] \int d^3\mathbf{k} \equiv [V/(2\pi^2)] \int_0^{\pi/d_z} dk_z \int_0^\infty k_\perp dk_\perp.
$
Note that pairing occurs only in the $\mathbf{k}$-space regions where both 
$E_{\mathbf{k},1}$ and $E_{\mathbf{k},2}$ have the same (positive) sign.
When $E_{\mathbf{k},1} \le 0$ and $E_{\mathbf{k},2} \ge 0$ or vice versa, 
the first term (on the right hand side) of Eq.~(\ref{eqn:op0}) inside 
the parentheses vanishes, reflecting that the pairing is not allowed
for those $\mathbf{k}$-space regions, and the quasiparticle and quasihole 
excitations are gapless.

In order to perform the integration over $k_\perp$ by hand, we need to find the
$\mathbf{k}$-space regions where the excitations are gapless, 
i.e. $E_{\mathbf{k},1} \le 0$ or $E_{\mathbf{k},2} \le 0$. 
The zeros of $E_{\mathbf{k},1}$ and $E_{\mathbf{k},2}$ are determined by
real and positive solutions of 
\begin{eqnarray}
0 = (1 - m_+^2/m_-^2) x^2 &+& 2(\xi_{k_z,+} - m_+ \xi_{k_z,-}/m_-) x \nonumber \\
&+& \xi_{k_z,\uparrow} \xi_{k_z,\downarrow} + |\Delta|^2,
\end{eqnarray}
where $x = k_\perp^2/(2m_+)$ and 
$
\xi_{k_z,\pm} = \epsilon_{k_z,\pm} - \mu_{\pm}.
$
Here, we introduced
$
\epsilon_{k_z,\pm} = (\epsilon_{k_z,\uparrow} \pm \epsilon_{k_z,\downarrow})/2,
$
where
$
\epsilon_{k_z,\sigma} = 2t_\sigma \left[1 - \cos(k_z d_z)\right]
$
is the energy dispersions in the $k_z$ direction.
Solutions of this equation ($x_<$ and $x_>$) depend
on $k_z$, and they give the locations of the zeros of $E_{\mathbf{k},1}$ 
and $E_{\mathbf{k},2}$ in the $k_\perp$-axis as a function of $k_z$. 
For instance, if both $x_<$ and $x_>$ are real and positive, 
then $E_{\mathbf{k},1}$ is gapless for $x_< \le k_\perp^2/(2m_+) \le x_>$ 
in some $k_z$ region $ z_{<,1}\le k_z \le z_{>,1}$, and $E_{\mathbf{k},2}$ 
is gapless for $x_< \le k_\perp^2/(2m_+) \le x_>$ in some other $k_z$ 
region $ z_{<,2} \le k_z \le  z_{>,2}$. If only $x_>$ is real and positive, 
$E_{\mathbf{k},1}$ is gapless for $0 \le k_\perp^2/(2m_+) \le x_>$ in some $k_z$ region 
$ z_{<,1} \le k_z \le z_{>,1}$, and $E_{\mathbf{k},2}$ is gapless for 
$0 \le k_\perp^2/(2m_+) \le x_>$ in some other $k_z$ region $ z_{<,2} \le k_z \le z_{>,2}      $.
If there is no real and positive solution, then the excitations are always gapped.

Given the gapless $\mathbf{k}$-space regions, Eq.~(\ref{eqn:op0}) can be 
written as
\begin{widetext}
\begin{eqnarray}
\frac{\pi}{a_{eff}} = 
\left( \int_{z_{<,1}}^{z_{>,1}} + \int_{z_{<,2}}^{z_{>,2}} \right) dk_z \ln\left[ 
\frac{x_> + \xi_{k_z,+} + \sqrt{(x_>+\xi_{k_z,+})^2 + |\Delta|^2}}
{x_< + \xi_{k_z,+} + \sqrt{(x_<+\xi_{k_z,+})^2 + |\Delta|^2}} \right]
-
\int_0^{\pi/d_z} dk_z \ln\left(
\frac{2\epsilon_{k_z,+}}{\xi_{k_z,+} + \sqrt{\xi_{k_z,+}^2 + |\Delta|^2}} \right),
\end{eqnarray}
where the first term on the right hand side is coming from the gapless, but
the second term is from the gapped $\mathbf{k}$-space regions.
Similarly, Eq.~(\ref{eqn:nup0}) can be written as
\begin{eqnarray}
\frac{4\pi^2 n_\uparrow}{m_+} =
&-& \left( \int_{z_{<,1}}^{ z_{>,1}} + \int_{z_{<,2}}^{ z_{>,2}} \right) dk_z 
\left[x_> - \sqrt{(x_>+\xi_{k_z,+})^2 + |\Delta|^2} - x_< + \sqrt{(x_<+\xi_{k_z,+})^2 + |\Delta|^2}\right] \nonumber \\
&+& 2 \int_{ z_{<,1}}^{ z_{>,1}} dk_z (x_> - x_<) 
+ \int_0^{\pi/d_z} dk_z \left( \sqrt{\xi_{k_z,+}^2 + |\Delta|^2} - \xi_{k_z,+} \right),
\end{eqnarray}
\end{widetext}
where again the first two terms on the right hand side is coming from the gapless, 
but the third term is from the gapped $\mathbf{k}$-space regions.
The density of $\downarrow$ fermions can be obtained by substituting 
$1 \to 2$ in the integration limits of the second term.
Our numerical calculations show that integrating $k_\perp$ by hand as 
described above and calculating the remaining $k_z$ integral numerically
is a much more stable approach compared to the one where both integrations 
are calculated numerically. In particular, this approach converges much 
faster than the latter on the BCS side, where integrations involve 
gapless $\mathbf{k}$-space regions.

\section{Normal-Superfluid Phase Boundary at $T = 0$}
\label{sec:app.b}

The phase boundary for the normal-superfluid transition can be found
from Eqs.~(\ref{eqn:op}),~(\ref{eqn:nup}) and~(\ref{eqn:ndo})
by setting $\Delta = 0$. Therefore, at the transition boundary, 
the self-consistency equations are uncoupled, i.e. the gap equation 
determines the critical effective scattering length, and the chemical 
potentials are determined by the number equations. 
At zero temperature, this leads to
\begin{eqnarray}
\frac{m_+ V}{4\pi a_{eff}} &=& \sum_{\mathbf{k}} \left[ 
\frac{1}{2\epsilon_{\mathbf{k},+}} 
- \frac{1 - \theta(-\xi_{\mathbf{k},\uparrow}) - \theta(-\xi_{\mathbf{k},\downarrow})}
{2\xi_{\mathbf{k},+}} \right],
\label{eqn:opSN} \\
N_{\sigma} &=& \sum_{\mathbf{k}} \theta(-\xi_{\mathbf{k},\sigma}).
\label{eqn:nsigSN}
\end{eqnarray}
Note that Eq.~(\ref{eqn:nsigSN}) is the number equation for noninteracting
fermions, and it is already solved in Sec.~\ref{sec:nil}. 
Similar to Appendix A, we can also simplify Eq.~(\ref{eqn:op}) by finding the 
zeros of $\xi_{\mathbf{k}, \sigma}$. This leads to
\begin{widetext}
\begin{eqnarray}
-\frac{\pi}{a_{eff}} =
\int_0^{k_{Fz,<}} dk_z \ln\left( \frac{\xi_{k_z,+}} 
{\xi_{k_z,+} - \frac{m_<}{m_+}\xi_{k_z,<}} \right)
+ \int_{0}^{k_{Fz,>}} dk_z \ln\left( \frac{\epsilon_{k_z,+}} 
{\xi_{k_z,+} - \frac{m_>}{m_+}\xi_{k_z,>}} \right) 
+ \int_{k_{Fz,>}}^{\pi/d_z} dk_z \ln\left( \frac{\epsilon_{k_z,+}} 
{\xi_{k_z,+}} \right),
\end{eqnarray}
\end{widetext}
where $k_{Fz,<} \equiv \min\{k_{Fz,\uparrow}, k_{Fz,\downarrow}\}$ and 
$k_{Fz,>} \equiv \max\{k_{Fz,\uparrow}, k_{Fz,\downarrow}\}$. For instance,
when $k_{Fz,<} \equiv k_{Fz,\downarrow}$, then $< \equiv \downarrow$ 
and $> \equiv \uparrow$ in the integrands.
Here, we again emphasize that integrating $k_\perp$ by hand as described 
above, and calculating the remaining $k_z$ integral numerically is a much 
more stable approach compared to the one where both integrations 
are calculated numerically.


\begin{thebibliography}{99}
\bibitem{nishidanote} In the weak coupling limit of 2D-3D mixtures, 
absence of the Cooper instability has already 
been pointed out in Ref.~\cite{nishida09}, where the ground state of 
the system is proposed to be an induced $p$-wave superfluidity of 
two-dimensional fermions, instead of being normal.
\bibitem{md} In cold-atom experiments, the peculiar momentum distribution 
of different topological phases would be smeared out by the trapping potential, 
but their marked signatures should still be present.
\bibitem{saddle} In mixed-dimensional systems, it is not obvious whether 
the saddle-point equations are sufficient to describe the BCS-BEC evolution 
of superfluidity or not. For instance, in a bilayer system~\cite{nishida10}, 
it has been proposed that the normal phase in the weak coupling region becomes an 
interlayer $s$-wave or intralayer $p$-wave superfluid phase depending on the layer
separation $d_z$. In addition, it is also possible that two two-dimensional 
atoms in different layers and one three-dimensional atom can form a 
trimer state whose binding energy scales as $~ 1/d_z^2$.
\end{thebibliography}
\end{document}